\documentclass[%
 reprint,
 amsmath,amssymb,
 aps,
]{revtex4-2}

\usepackage{graphicx}
\usepackage{dcolumn}
\usepackage{bm}
\usepackage{xspace}
\usepackage{setspace}
\usepackage{subfig}
\usepackage{verbatim}\usepackage{booktabs} 
\usepackage{siunitx}
\usepackage{amsmath}
\usepackage{ragged2e}
\usepackage{soul}
\usepackage{afterpage}

\begin{document}

\preprint{APS/123-QED}

\title{Nonreciprocal Transport in chiral Mo$_3$Al$_2$C Near the Superconducting–Normal Transition}

\author{Jeongsoo Park$^1$, Sang-Wook Cheong$^2$, and Xianghan Xu$^{1}$}
\thanks{xu001395@umn.edu}
\affiliation{$^1$University of Minnesota -- Twin Cities, Minneapolis, Minnesota 55455, USA \\ $^2$Rutgers University, Piscataway, New Jersey 08854, USA}

\date{\today}

\begin{abstract}
Mo$_3$Al$_2$C is an intriguing material that simultaneously hosts crystallographic chirality, an electronic instability that further distorts the lattice below around 155 K, and a superconducting transition near 8 K. We investigate nonreciprocal electrical transport in bulk single-crystalline Mo$_3$Al$_2$C using an AC probe over wide ranges of temperature, current, and magnetic field, with various field orientations relative to the current, leading to a comprehensive construction of the first-harmonic ($R^{1\omega}$) and second-harmonic ($R^{2\omega}$) resistance responses. Remarkably, a clear $R^{2\omega}$ signal, the hallmark of nonreciprocal transport, is found near the boundary of the normal and superconducting phases in the phase diagram. Phenomenologically, this effect arises from direction-dependent constructive or destructive coupling between the current-induced intrinsic magnetization in the chiral lattice and the externally applied field parallel to the current. For perpendicular field orientations, the persistent $R^{2\omega}$ response suggests a toroidal-induced nonreciprocal effect associated with the polarization arising from the charge-density-wave phase and the orthogonal applied field H. These results demonstrate that intrinsic bulk Mo$_3$Al$_2$C provides an exceptional platform for tunable nonreciprocal transport rooted in chirality and polarity, offering new opportunities for superconducting quantum technologies.

\end{abstract}

\maketitle

\section{Introduction}
Nonreciprocal transport, the asymmetry of electrical response with respect to current or field reversal, has emerged as a defining signature of systems lacking both inversion and time-reversal symmetry~\cite{Tokura2018}. In such systems, the conventional Ohmic relation $V(I)=-V(-I)$ breaks down, giving rise to intrinsic direction-dependent resistances and rectification effects~\cite{daido2022,Ideue2020}. Experimentally, nonreciprocal transport has been observed in a growing range of materials, from superconductors and chiral magnets to polar semiconductors and van der Waals heterostructures~\cite{Choe2019,ando2020,aoki2019,ideue2017,Holmes2020,1Wu2022}. The sensitivity of the effect to lattice symmetry, domain configuration, and spin texture makes it a versatile probe of coupled spin–lattice–charge degrees of freedom~\cite{Ergecen2023,Nakamura2025}. Importantly, its tunability through temperature, magnetic field, and electrostatic gating opens pathways for controlling directional charge flow, a critical ingredient for rectifiers, neuromorphic devices, and energy-efficient spintronics~\cite{wu2022,Wakamura2024,Xiong2024}.

Superconductors, in particular, provide an exceptionally rich platform for realizing nonreciprocal transport. In noncentrosymmetric superconductors, the absence of inversion symmetry allows the supercurrent to couple asymmetrically to the underlying electronic and momentum structure~\cite{carnicom2018}. When time-reversal symmetry is further broken, for instance, by an external magnetic field, the condensate acquires a preferred direction of motion, leading to direction-dependent transport responses, i.e., the so-called superconducting diode effect~\cite{nagata2025,kealhofer2023,gupta2023}. This effect involves microscopic processes such as asymmetric vortex dynamics, underscoring how broken symmetries can govern dissipation and transport, and is directly relevant for implementing nonreciprocal elements in superconducting quantum technologies. Experimentally, nonreciprocal transport has been reported in a wide range of superconducting systems, including low-dimensional materials such as MoS$_2$ and NbSe$_2$~\cite{wakatsuki2017,zhang2020}, artificial superlattices~\cite{ando2020}, polar oxides like SrTiO$_3$~\cite{itahashi2020}, and twisted cuprate superconductor devices ~\cite{2Wang2025}. More recently, a pronounced nonreciprocal response has also been observed in CsV$_3$Sb$_5$, whose centrosymmetric lattice nonetheless hosts superconducting states with spontaneously inversion symmetry breaking~\cite{wu2022}. These findings highlight that nonreciprocity in superconductors can originate not only from structural asymmetry but also from electronically driven symmetry breaking, offering new routes to explore chiral superconductivity, correlated quantum transport, and underlying topology.

Recently, single crystals of the noncentrosymmetric superconductor Mo$_3$Al$_2$C ($T_c$ near 8 K) have been discovered to host a chiral lattice structure at room temperature~\cite{wu2024}. Upon cooling, a charge density wave transition near 155 K further distorts the lattice into an \textit{R}3 chiral and polar phase, introducing multiple intertwined symmetry-breaking characters~\cite{wu2024}. These features make Mo$_3$Al$_2$C a particularly promising platform for investigating nonreciprocal transport in a bulk superconducting system. In this work, we perform comprehensive AC transport measurements to investigate the nonreciprocal response of Mo$_3$Al$_2$C using the second-harmonic resistance ($R^{2\omega}$) as a quantitative indicator. A pronounced enhancement of the nonreciprocal signal appears near the superconducting-normal phase boundary, with rich dependencies on the relative orientations of the current and magnetic field, arising from the coexistence and synergetic interplay of chiral and polar components with the applied field. In contrast to most prior studies on microfabricated thin-film devices~\cite{yasuda2019,zhang2025,yoshimi2022}, our observation in bulk single crystals shows that nonreciprocal transport can emerge as an intrinsic property of the material itself, free from device-fabrication complexity.
\section{Experiment}
The crystal used in this study was synthesized following the procedure described in Ref.~\cite{wu2024}.The crystal contains left- and right-handed chiral domains, but with a strongly imbalanced population (roughly 90:10), as determined by single-crystal x-ray diffraction, while electron microscopy reveals evenly distributed polar domains without preferred orientation below the charge-density-wave transition near 155 K. The sample was oriented by Laue diffraction so that the electrical current flows along the cubic [110] direction, with the surface normal aligned along [1-11]. Four-probe contacts were made using silver epoxy, with AC current applied between probes 1 and 4 and the voltage measured between probes 2 and 3 using a lock-in amplifier. [Fig. 1(c)]. \\
\begin{figure*}[]
    {\includegraphics[width=0.8\linewidth]{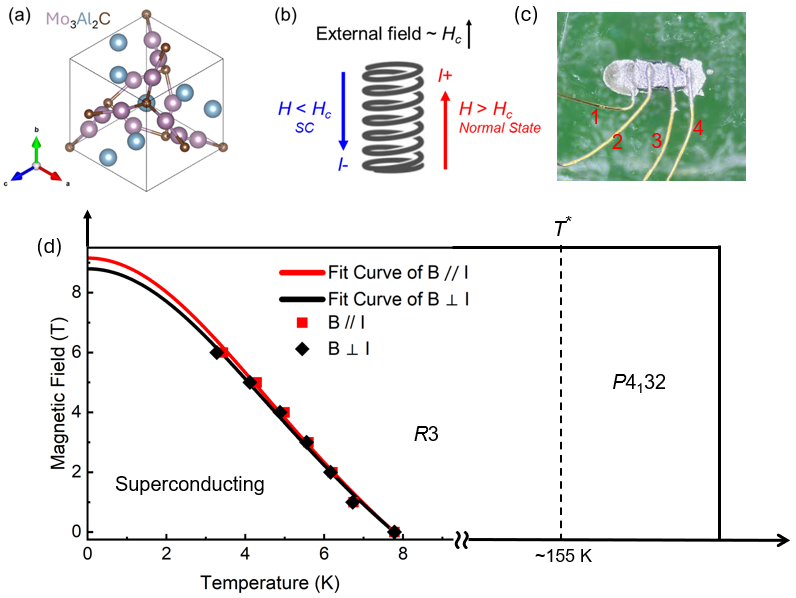}}
    
    \caption{\justifying (a) Cubic structure (P4$_1$32) of Mo$_3$Al$_2$C above $T^*=155$ K where CDW happens (b) A diagram of the chiral structure of Mo$_3$Al$_2$C. The chiral current induces magnetic field which enhances or diminishes the external magnetic field, switching from the normal to superconducting state, vice versa. (c) A circuit image of Mo$_3$Al$_2$C. The number on the wire denotes the probe number. (d) A M-T phase diagram of Mo$_3$Al$_2$C. $T^*=155$ K denotes the temperature where CDW happens.}
    \label{}%
\end{figure*}
\indent The experiment began by measuring the first-harmonic ($R^{1\omega}$) and second-harmonic ($R^{2\omega}$) resistances as the temperature was swept from 3.2 K to 15 K with 10 mA of AC current, and applied magnetic fields parallel and perpendicular to the current up to ±9 T. 
\begin{figure*}[t]
    \centering
    {\includegraphics[width=1\linewidth]{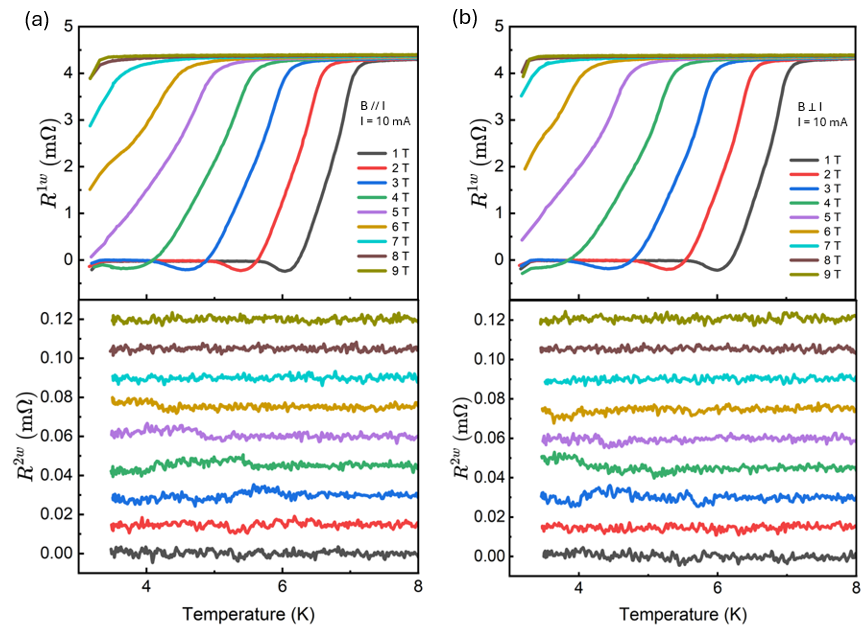}}
    \caption{\justifying Nonreciprocal magnetoresistance of (sample dimension). (a), (b) Plots of first (top) and second (bottom) harmonic resistances when the magnetic field is parallel and perpendicular to the current, respectively. The dashed line is drawn at the temperature where $R^{1\omega}$ drops by 50\% for 3 T, comparing the transition temperature in $R^{1\omega}$ and $R^{2\omega}$. The $R^{2\omega}$ signals in both (a) and (b) are offset by $0.015$ m$\Omega$.}
    \label{fig:example}%
\end{figure*}
Next, the temperature was held at 5 K, and $R^{2\omega}$ was measured with currents ranging from 5 mA to 20 mA in steps of 2.5 mA.
\begin{figure*}[t]
    \centering
    {\includegraphics[width=1\linewidth]{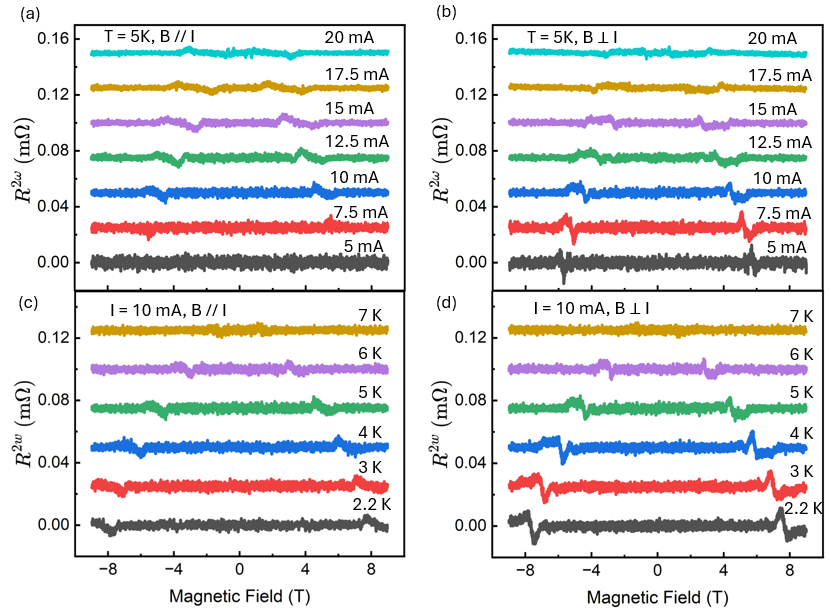}}
    \caption{\justifying Second harmonic resistance ($ R^{2\omega} $) as a function of magnetic field for different currents (a, b) and temperatures (c, d), with the field parallel and perpendicular to the current. The $R^{2\omega}$ signals in all panels are offset by $0.015$ m$\Omega$.}
    \label{fig:example}%
\end{figure*}
Finally, with the current fixed at 10 mA, $R^{2\omega}$ was measured at temperatures ranging from 2.2 K to 7 K [Fig. 3(c, d)]. The maximums of the absolute value of $R^{2\omega}$ were determined relative to their baselines to assess the nonreciprocal responsiveness of transport [Fig. 4], with uncertainties calculated as the standard deviation of each baseline. Additionally, $R^{2\omega}$ was measured as the angle between the magnetic field and current varied, and plotted in polar coordinate [Fig. S1]. All plotted $R^{2\omega}$ data were antisymmetrized with respect to the magnetic field to extract the nonreciprocal transport component, which is expected to be odd under field reversal.
\section{Results and Discussion}
Generally, applying a magnetic field to a chiral material breaks both time-reversal and spatial-inversion symmetries, enabling nonreciprocal transport through the magnetochiral effect without requiring superconductivity ~\cite{rikken2001,Nakamura2025}. Nevertheless, the nonreciprocal response is expected to be strongly enhanced near the superconducting–normal transition for the following reason. From a symmetry perspective, driving a current through a chiral texture induces an effective magnetic field that can either reinforce or oppose the external field, depending on whether the current and field are parallel or antiparallel. When the external field is tuned close to the critical field \textit{H$_c$}, this constructive or destructive interference effectively pushes the system into or out of the superconducting state, amplifying the resistivity difference and resulting in a much larger $R^{2\omega}$ signal, schematically illustrated in Fig. 1b. Another possible root cause for nonreciprocal transport is the so-called toroidal effect, i.e., when there is a magnetic field perpendicular to the direction of electric polarization, nonreciprocal tranport happens along the third orthogonal direction. Symmetry-wise, the toroidal effect is equivalent to a effective intrinsic velocity vector propagating parallel to the curl of polarization and magnetization, causing constructive and destructive coupling to the observed propagation~\cite{cheong2018,Yamaguchi2025}. \\
\indent To experimentally validate the hypothesis described above, first, a superconducting \textit{H-T} phase diagram (Fig. 1d) is established based on the $R^{1\omega}$ data displayed in Fig. 2a-2b, with field parallel and perpendicular to the current, respectively. The superconducting transition temperature \textit{T$_c$} is defined as the temperature where $R^{1\omega}$ drops by 50\%. The upper critical field \textit{H$_{c2}$(T)} were fitted with the generalized Ginzburg-Landau model~\cite{tinkham2004}:
\begin{equation}
    H_{c2}(T) = H_{c2}(0)\cdot\frac{1-t^2}{1+t^2}
\end{equation}
where $t=T/T_c$. The material is found to have a zero-field critical temperature $T_c = 7.78$ K. The fit parameter $H_{c2}(0)$ is $9.15\pm0.11$ T and $8.79\pm0.10$ T for the parallel and perpendicular cases, respectively, suggesting a small but detectable gap anisotropy  ~\cite{tinkham2004,tsurkan2011}. The transition temperatures at 7 T, 8 T, and 9 T could not be estimated, as they fall outside the measured temperature range. \\
\indent The lower sub-panels of Fig. 2 display temperature-dependent $R^{2\omega}$ at various field, revealing non-zero signals primarily near these transition regions for fields of 3 T to 6 T, while higher fields (7-9 T) show negligible responses as their transitions fall below the measured temperature range, or the superconducting transition is completely suppressed. The reason why $R^{2\omega}$ signals are weak at 1 T and 2 T is that the magnitude of the field is small, because of $R^{2\omega} \propto \vec{H}\cdot\vec{I}$ for the magnetochiral mechanism. The $R^{2\omega}$ anomalies are also present when the magnetic field is applied perpendicular to the current, corresponding to a nonreciprocity from the toroidal effect~\cite{cheong2018,Yamaguchi2025}. However, their correspondence with the transition is noticeably weaker than the parallel situation, likely because the response is significantly broadened by the presence of polar domains oriented along the eight possible body-diagonal directions of the cubic high-temperature phase framework~\cite{wu2024}.\\
\indent Fig. 3a-3b presents $R^{2\omega}$ as a function of the magnetic field at a fixed temperature of 5 K, for current amplitudes ranging from 5 mA to 20 mA in 2.5 mA increments. In Fig. 3a, where the magnetic field is parallel to the current (B $\|$ I), the curves exhibit pronounced peaks near the superconducting-normal transitions, from approximately 5.53 T at 7.5 mA to around 3.05 T at 20 mA. In Fig. 3b, the perpendicular orientation (B $\perp$ I) shows similar anomalies near the transitions, while the $R^{2\omega}$ signals exhibit more complicated peak splitting and positive–negative asymmetries. These features are again likely attributable to the presence of multiple polar domains. In both field orientations, the peak positions shift toward lower fields as the current increases, consistent with the picture that a larger current suppresses the superconducting critical field. Meanwhile, the peak intensities evolve non-monotonically, as discussed further in Fig. 4. Fig. 3c-3d presents the $R^{2\omega}$ response as a function of magnetic field at a fixed current of 10 mA, measured over temperatures from 2.2 K to 7 K for the configurations with magnetic field parallel to the current (B $\|$ I) and perpendicular to the current (B $\perp$ I), respectively. In both geometries, pronounced $R^{2\omega}$ anomalies emerge near the superconducting–normal transition, occurring at field–temperature boundaries consistent with the phase diagram in Fig. 1d. Overall, the temperature and current evolution of these features reinforces the interpretation that the nonreciprocal response is strongly enhanced in the vicinity of the transition, where the interplay between superconductivity, vortex dynamics, chirality, and polarity becomes most significant.\\
\begin{figure*}[]
    \centering
    {\includegraphics[width=1\linewidth]{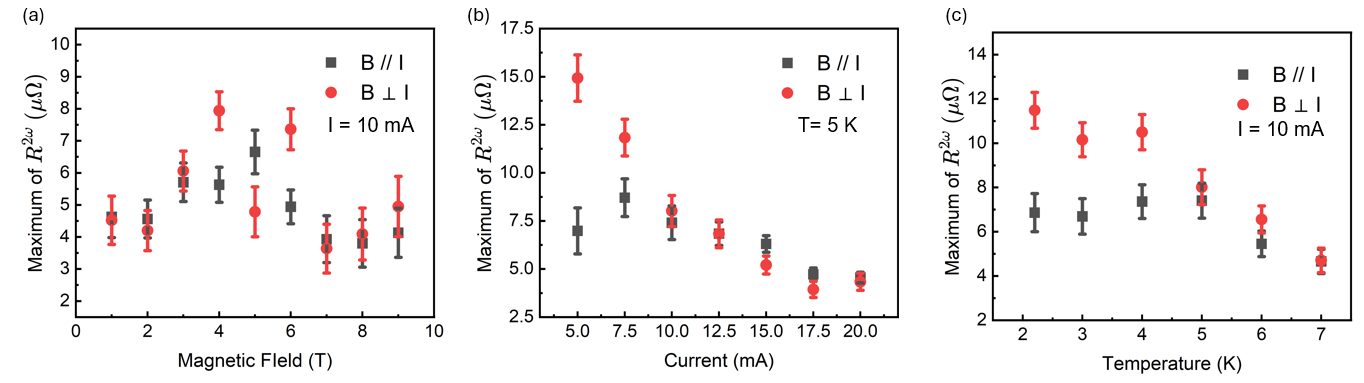}}
    \caption{\justifying (a, b, c) Maximum value of $R^{2\omega}$ vs. field, current and temperature from Fig. 2a, 2b, 3a, 3b, 3c, and 3d, respectively. The uncertainty of each data point was determined to be the standard deviation of signal's noise level.}
    \label{fig:example}%
\end{figure*}
\indent Fig. 4 quantitatively examines how external parameters influence the magnitude of the nonreciprocal response in Mo$_3$Al$_2$C, enabling us to identify the optimal magnetic field, current, and temperature conditions that maximize $R^{2\omega}$, by compiling the peak values of the antisymmetrized $R^{2\omega}$ extracted from Fig. 2a, 2b,3a,3b,3c,and 3d, respectively. In Fig. 4a, the maximum $R^{2\omega}$ versus magnetic field (1--9 T) for both $B \parallel I$ and $B \perp I$ decreases beyond approximately 5 T. Fig. 4b shows the peak $R^{2\omega}$ versus current at 5 K, with a noticeable drop at large current, beyond roughly 7.5 mA for the parallel case and 5 mA for the perpendicular case. Fig. 4c reveals a similar reduction with increasing temperature for a fixed 10 mA.\\
\indent The overall non-monotonic behavior originates from two competing mechanisms. On one hand, the relation $R^{2\omega} \propto \vec{H}\cdot\vec{I}$ suggests that larger current and field should enhance the nonreciprocal response. On the other hand, stronger magnetic fields suppress superconductivity and weaken vortex pinning, thereby reducing the signal. The interplay between these reinforcing and competing tendencies naturally leads to a dome-shaped dependence of $R^{2\omega}$ on both field and current, with a maximum at intermediate values. The only exception is the current dependence in the $B \perp I$ configuration, where the monotonic decrease at high current may arise from current-induced sliding of the charge-density-wave state, which weakens the polarization and consequently suppresses the nonreciprocal response. Together, the trends in Fig. 4 demonstrate that nonreciprocal transport in Mo$_3$Al$_2$C is optimized within an intermediate window, providing a quantitative guide for tuning external conditions.\\
\indent Additionally, the heat capacity over temperature $(C_p /T)$ was measured and plotted [Fig. S2], revealing the thermodynamic signatures of its superconducting and charge density wave (CDW) phases. A sharp anomaly at $ T_c \approx 8 $ K marks the superconducting transition. The dashed line represents a fit to the normal-state heat capacity, $ C_p / T = \gamma + \beta T^2 $, yielding $ \gamma = 0.0222(7) $ J mol$^{-1}$ K$^{-2}$ and $ \beta = 2.8069(10) \times 10^{-4} $ J mol$^{-1}$ K$^{-4}$. The obtained normalized jump $ \Delta C_p / (\gamma T_c) = 2.15 \pm 0.13$,  exceeding the weak-coupling BCS limit of 1.43 and indicating strong electron-phonon coupling in the superconducting state, which also agrees with prior research done by A. B. Karki et al~\cite{karki2010} on a polycrystalline sample. The main possible uncertainty originates from the mass measurement of the sample, which is $(1.70\pm0.05)$ mg.  This indicates a strongly coupled superconducting state, likely mediated by enhanced electron–phonon and spin–orbit interactions. Such strong coupling, together with the noncentrosymmetric crystal structure of Mo$_3$Al$_2$C, can give rise to mixed-parity pairing and pronounced momentum dependence, which provide a possible microscopic origin for the observed nonreciprocal transport behavior. A feature in the heat capacity near 155 K corresponds to the CDW transition, distorting the cubic lattice into a rhombohedral structure and creates the interesting polar metal state via the trilinear coupling, which enrich the nonreciprocal responses through a toroidal effect as presented above.
\section{Conclusion}
This study investigates the nonreciprocal transport properties of Mo$_3$Al$_2$C, where the chiral lattice and polar charge-density-wave phase break inversion symmetry, and an applied magnetic field breaks time reversal symmetry, thereby permitting a finite second-harmonic transport response. Using second-harmonic AC probe, we observe pronounced $R^{2\omega}$ signals near the normal-superconducting transition, where superconductivity transition significantly amplifies the nonreciprocity. These findings not only establish Mo$_3$Al$_2$C as a compelling platform for studying superconducting diode effects, but also quantitatively explore the external stimuli under which the nonreciprocal response is maximized. \\
\indent The toroidal response in Mo$_3$Al$_2$C is particularly intriguing given the presence of uniformly distributed polar domains. In a scenario with equal populations of oppositely oriented polar domains, one might anticipate the $R^{2\omega}$ signal for magnetic fields perpendicular to the current to cancel out. The persistence of a finite response therefore suggests that the opposite polar domains may host different critical fields, resulting in an incomplete cancellation. Consistent with this picture, Fig. 3b and Fig. 3d exhibit a characteristic positive–negative two-peak structure in the $R^{2\omega}$ anomaly. Looking ahead, microscopy techniques with sufficient spatial resolution to resolve the field-dependent response of individual polar domains would provide desirable insight. \\
\indent While most studies of nonreciprocal transport in superconductors rely on nano-fabricated thin-film devices or engineered junctions to isolate symmetry-breaking effects~\cite{yasuda2019,zhang2025}, our measurements on bulk single crystals of Mo$_3$Al$_2$C preserve the material's structural integrity, enabling broader exploration of intrinsic magnetochiral effects in the pristine chiral lattice. Recently, Orban et al. demonstrated the superconducting diode effect in Josephson junctions formed by stacking bulk Mo$_3$Al$_2$C single crystals of the same or opposite handedness, using DC transport to quantify asymmetries in the critical currents \textit{$I_{c+}$} and \textit{$I_{c-}$} for opposite current directions~\cite{orban2025}. Their results further illustrate how this chiral superconductor can serve as versatile building blocks for functional quantum devices, and our study complements these findings by revealing second-harmonic responses under AC excitation, underscoring the material's broad potential for symmetry-driven quantum transport.\\
\indent Spin–orbit coupling (SOC) could also enhance nonreciprocal transport by coupling the direction of charge flow to the material’s chiral lattice, leading to asymmetric scattering for ±k propagations ~\cite{Putilov2024}. Therefore, SOC may act as an amplifying factor for the observed nonreciprocity, and investigating its role more directly would help clarify its contribution and identify practical routes to further boost the effect. A promising direction is to examine isostructural superconductors in this family that incorporate heavier elements with stronger SOC effect ~\cite{Ying2019, Yang2025}.
\bibliography{bib}
\newcommand\blankpage{%
    \null
    \thispagestyle{empty}%
    \addtocounter{page}{-1}%
    \newpage}
\begin{figure*}[]
    {\includegraphics[width=1\linewidth]{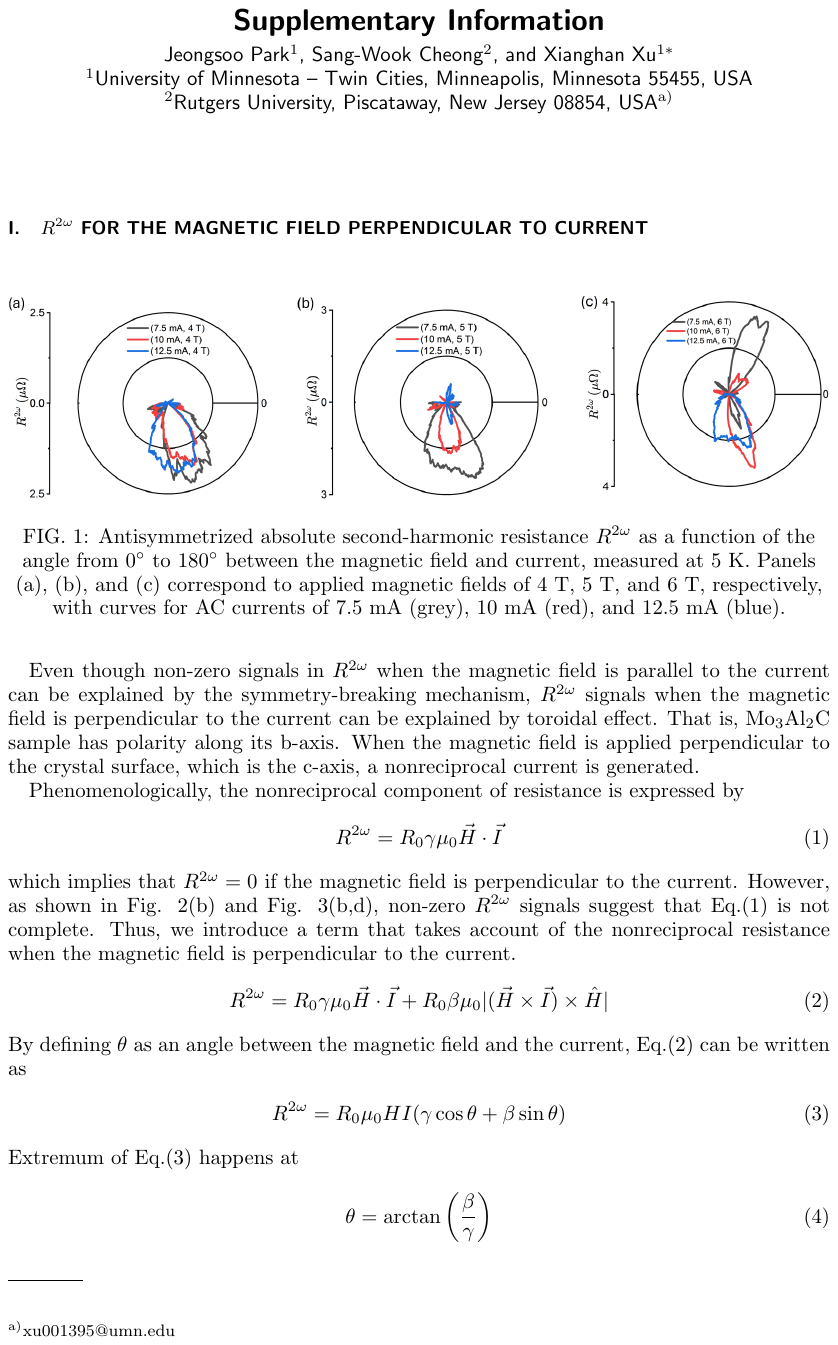}}
    \caption*{}
    \label{}
\end{figure*}
\begin{figure*}[]
    {\includegraphics[width=1\linewidth]{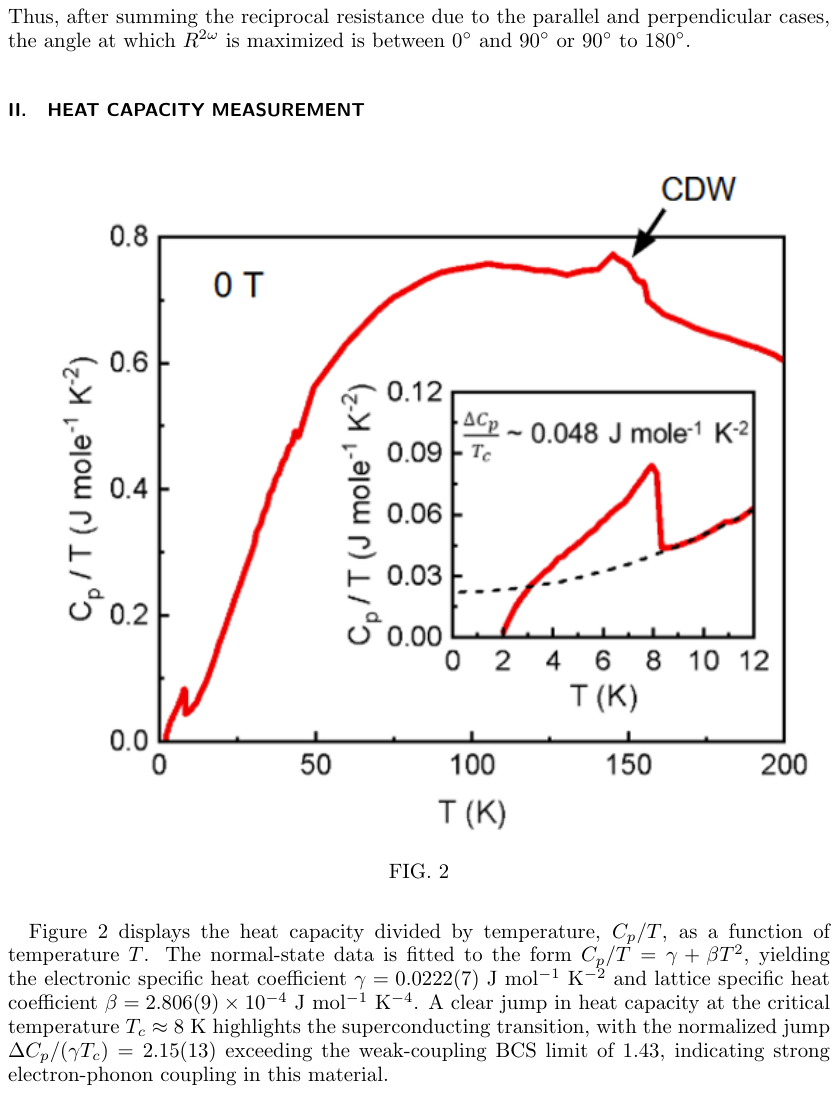}}
    \caption*{}
    \label{}
\end{figure*}
\end{document}